\documentclass[reprint,aps,pra,superscriptaddress,notitlepage]{revtex4-1}
\usepackage{amssymb,amsmath}
\usepackage{graphicx}
\usepackage{color}
\usepackage[english]{babel}
\usepackage[caption=false]{subfig}
\usepackage[colorlinks=true,allcolors=blue]{hyperref}
\usepackage{xcolor}
\usepackage{makecell}
\usepackage[T1]{fontenc}

\begin{document}

\def\simlt{\mathrel{\lower .3ex \rlap{$\sim$}\raise .5ex \hbox{$<$}}}

\title{\textbf{\fontfamily{phv}\selectfont 
Fabrication process and failure analysis for robust quantum dots in silicon}}
\author{J. P. Dodson}
\affiliation{Department of Physics, University of Wisconsin-Madison, Madison, WI 53706, USA}
\author{Nathan Holman}
\affiliation{Department of Physics, University of Wisconsin-Madison, Madison, WI 53706, USA}
\author{Brandur Thorgrimsson}
\affiliation{Department of Physics, University of Wisconsin-Madison, Madison, WI 53706, USA}
\author{Samuel F. Neyens}
\affiliation{Department of Physics, University of Wisconsin-Madison, Madison, WI 53706, USA}
\author{E. R. MacQuarrie}
\affiliation{Department of Physics, University of Wisconsin-Madison, Madison, WI 53706, USA}
\author{Thomas McJunkin}
\affiliation{Department of Physics, University of Wisconsin-Madison, Madison, WI 53706, USA}
\author{Ryan H. Foote}
\affiliation{Department of Physics, University of Wisconsin-Madison, Madison, WI 53706, USA}
\author{L. F. Edge}
\affiliation{HRL Laboratories, LLC, 3011 Malibu Canyon Road, Malibu, CA 90265, USA}
\author{S. N. Coppersmith}
\affiliation{Department of Physics, University of Wisconsin-Madison, Madison, WI 53706, USA}
\affiliation{University of New South Wales, Sydney, Australia}|
\author{M. A. Eriksson}
\affiliation{Department of Physics, University of Wisconsin-Madison, Madison, WI 53706, USA}

\begin{abstract}
We present an improved fabrication process for overlapping aluminum gate quantum dot devices on Si/SiGe heterostructures that incorporates low-temperature inter-gate oxidation, thermal annealing of gate oxide, on-chip electrostatic discharge (ESD) protection, and an optimized interconnect process for thermal budget considerations. This process reduces gate-to-gate leakage, damage from ESD, dewetting of aluminum, and formation of undesired alloys in device interconnects. Additionally, cross-sectional scanning transmission electron microscopy (STEM) images elucidate gate electrode morphology in the active region as device geometry is varied. We show that overlapping aluminum gate layers homogeneously conform to the topology beneath them, independent of gate geometry, and identify critical dimensions in the gate geometry where pattern transfer becomes non-ideal, causing device failure.
\end{abstract}

\maketitle

\section{Introduction}
Developing a suitable physical system for quantum computation has received much attention in the past two decades. Since Loss and Divencenzo's proposal \cite{Loss:1998p120}, significant progress has been made using spins in solid-state systems as quantum bits (qubits). Coherent control of semiconductor quantum dots using spin degrees of freedom was first demonstrated in GaAs/Al$_{0.3}$Ga$_{0.7}$As heterostructures \cite{Petta:2005p2180, Koppens:2006p766}. This particular heterostructure found initial success due to the small electron effective mass in GaAs and depletion mode operation of devices. This allows for large, single-layer gate geometries to tune devices into the few-electron regime \cite{Ciorga:2000p16315}. Although fabrication and characterization of one-qubit devices in GaAs has become routine \cite{Petta:2005p2180, Koppens:2006p766, Petta:2004p1586, PioroLadriere:2008p776, Laird:2010p075403}, short coherence times of spin-qubits due to the presence of nuclear spins \cite{Coish:2004p5340} make it difficult to achieve fidelities necessary for fault-tolerant operation \cite{Fowler:2012p032324}.
\begin{figure*}
\includegraphics[width=1.0\textwidth]{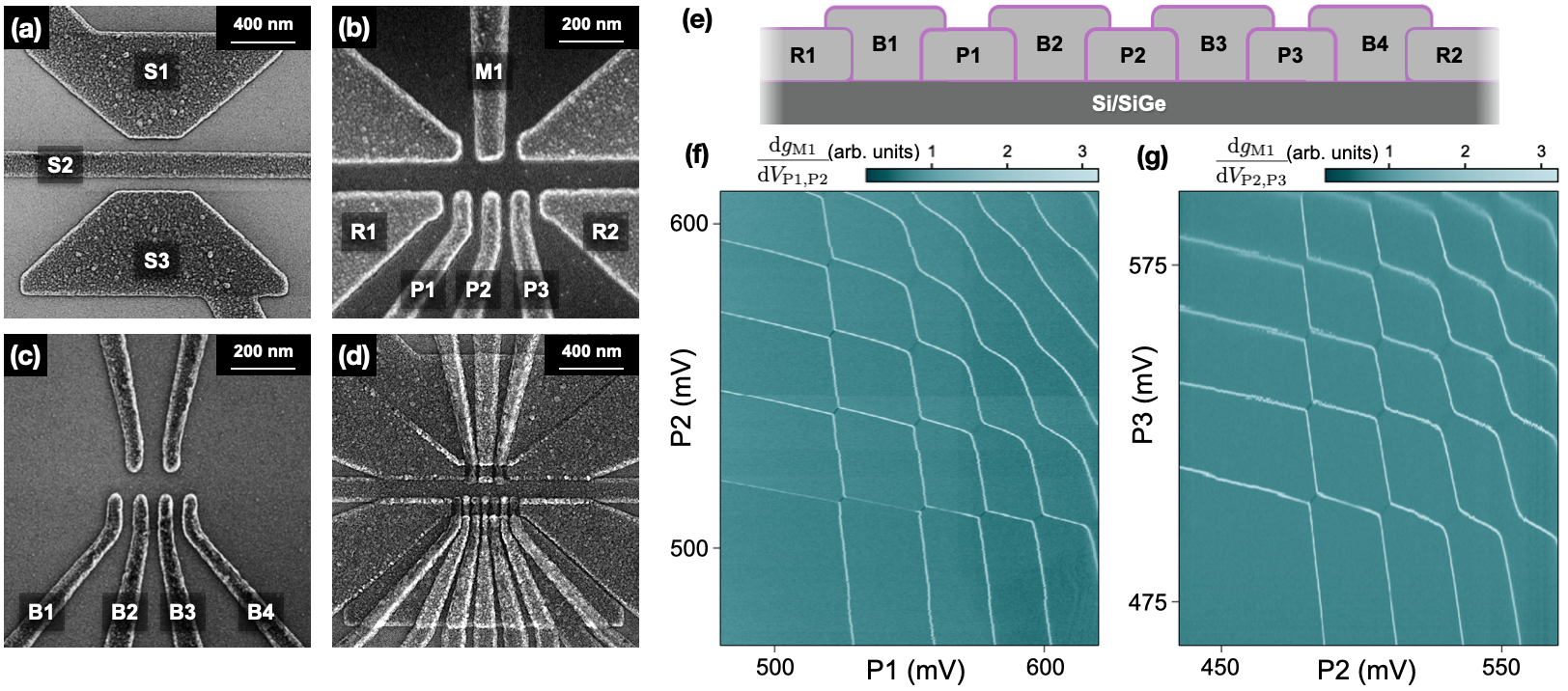}
\caption{\label{fig:fig1} Scanning electron microscope images of a triple-dot device active region and stability diagrams. (a) Screening gates are patterned in the first layer. (b) The accumulation gate layer patterns reservoirs and plunger gates. (c) Barrier gates are patterned on the third layer to fill gaps between plunger gates. (d) All three layers of a completed triple-dot device. (e) Cross-sectional schematic of the gate stack in the quantum dot channel between gates S2 and S3, where the light purple border around the Al electrodes is native AlO$_x$. The screening gates from (a) are not shown because they border the quantum dot channel. (f, g) Stability diagrams of plunger gate pairs P1/P2 and P2/P3 where the differential conductance $dg_{\text{M1}}/dV_P$ beneath charge sensing dot M1 is measured by modulating the voltage on both plunger gates simultaneously, demonstrating electrostatic control of the few-electron regime in each quantum dot pair.}
\end{figure*}
\\ \indent Silicon-based approaches have significantly improved qubit performance in solid-state spin systems in part due to the spinless nucleus of $^{28}$Si. Experiments using isotopically purified $^{28}$Si have shown average single-qubit control fidelity in excess of 99.9\% \cite{Yoneda:2018p102}. However, in Si, the lithographic demands are more stringent because of the larger electron effective mass. Different gate designs have been explored, including open geometries, which use global top gate(s) and several depletion gates to form each quantum dot \cite{Borselli:2011p063109, Wu:2014p11938} and tight geometries which use linear, overlapping gates with dedicated accumulation and depletion electrodes for each quantum dot \cite{Angus:2007p845, Zajac:2016p054013}. While both general designs have generated two-qubit devices in Si/SiGe \cite{Zajac:2018p439, Watson:2018p633, Xue:2019p21011} and Si-MOS \cite{Veldhorst:2015p410, Huang:2019p7757}, the overlapping gate architecture has clear advantages in scaling to larger systems. A 9-dot array has already been demonstrated \cite{Zajac:2016p054013} and similar architectures can be scaled into much larger arrays. Additionally, quantum dots form in predictable locations with tunnel couplings that can be well-controlled using a single gate \cite{Zajac:2015p223507, Neyens:2019p08216}. While there are benefits in choosing a linear, overlapping gate architecture, developing a high yield fabrication process is challenging.
\\ \indent In this article, we investigate many yield limiting steps in the fabrication of linear, overlapping aluminum gate quantum dot devices, showing failure analysis of critical interfaces and providing improved process steps for difficulties experienced in typical fabrication procedures. We present results on three main topics: low-temperature oxidation of inter-gate aluminum oxide (AlO$_x$), cross-sectional scanning transmission electron microscopy (STEM) analysis of overlapping Al gate geometries, and characterization of interconnects between the device bond pads and active region---the local region surrounding quantum dots. Four low-temperature oxidation techniques are compared for enhancement of the native AlO$_x$ that electrically isolates subsequent gate electrode layers from each other. STEM analysis investigates test structures with varying dot-to-dot pitch, characterizing how different gate geometries affect gate electrode morphology and the filling of barrier gates in gaps between plunger gates. For the interconnects, we optimize the process flow to allow for thermal annealing of the Al$_2$O$_3$ grown by atomic layer deposition (ALD), incorporation of the low-temperature oxidation techniques presented, and integration of on-chip electrostatic discharge (ESD) protection.

\section{Fabrication Methods}

\indent Layer-by-layer fabrication of overlapping Al gate quantum dot devices is shown in Figure~\ref{fig:fig1}. A top down scanning electron microscope (SEM) image of each Al gate layer is shown in Figure~\ref{fig:fig1}(a-c) and a completed device active region with all three layers is shown in Figure~\ref{fig:fig1}(d). The device consists of three quantum dots with four barrier gates on the bottom side and a single integrated charge sensing dot on the top side. This particular geometry is shown because it is a unit cell for an architecture that can be linearly tiled \cite{Zajac:2016p054013}. All layers are patterned using an Elionix ELS G-100 electron beam writer on PMMA 495 A4 resist. Layers are aligned using 200 nm thick evaporated gold alignment marks, achieving better than 10 nm overlay accuracy on average. The Al gate metal is deposited using a Lesker PVD 75 electron-beam evaporator at 0.3 \AA/s at a pressure of $\sim$9$\times 10^{-7}$ Torr. All electron-beam lithography (EBL) steps using a PMMA resist stack are developed in 3:1 IPA:MIBK and lifted off using Remover PG heated to 75$^\circ$C for 1 hour, followed by a solvent clean. After lift-off, the Al gates on the first two layers are further oxidized using a plasma ash technique that will be described in detail below. A more thorough fabrication process is described in the Appendix.
\\ \indent Each set of gates takes on a specific role in manipulating the chemical potential of the two-dimensional electron gas (2DEG). The first layer, shown in Figure~\ref{fig:fig1}(a), acts to screen stray electromagnetic fields from the second and third layers. The screening gates are used to deplete regions beneath them, defining a long, thin channel in the 2DEG that can be accumulated/depleted by gates from subsequent layers. A 150 nm wide central screening gate is patterned to prevent current injection from the top side to the bottom side. The second layer, shown in Figure~\ref{fig:fig1}(b), consists of accumulation gates, including reservoir gates for Fermi level control of electron reservoirs, and plunger gates for tuning electron occupation within quantum dots. Figure~\ref{fig:fig1}(c) shows the third layer, which patterns barrier gates designed to deplete the 2DEG and tune the tunnel coupling into and out of the quantum dots. The final device with all three layers is shown in Figure~\ref{fig:fig1}(d). The width of all plunger/barrier gates is increased to 100 nm  before crossing the back edge of the screening gates in order to minimize step coverage failure. In this particular geometry, plunger gates are nominally 70 nm wide with a 120 nm pitch, and barrier gates are 60 nm wide, filling a 50 nm gap. Figure~\ref{fig:fig1}(e) shows a cross-sectional schematic of the quantum dot channel in Figure~\ref{fig:fig1}(d), where screening gates S2 and S3 border the channel. The device is cooled in a dilution refrigerator with a base temperature of $<$50 mK and electron temperature $T_e = 100 \pm 10$ mK. Figure~\ref{fig:fig1}(f, g) show charge stability diagrams in the few-electron regime between each pair of adjacent quantum dots P1/P2 and P2/P3, respectively, as measured by an integrated charge sensing dot beneath M1. M1 is biased into a configuration such that it is on the highest sloped region of a Coulomb blockade peak, where changes in charge occupancy of quantum dots P1-P3 result in detectable shifts in current through the sensor dot \cite{Field:1993p1477}. The differential conductance $dg_{\text{M1}}/dV_P$ is measured using standard lock-in techniques as detailed in Ref.~\cite{Elzerman:2003p728}, where the voltage on both plunger gates are modulated simultaneously. High resolution plots were taken over a 24 hour period to demonstrate stability in the few-electron regime.

\section{Low-Temperature Inter-gate Oxidation}
\indent Gate layers are electrically isolated via the AlO$_x$ that grows natively on the Al gate electrodes. This allows for the omission of blanket dielectric films such as ALD-grown Al$_2$O$_3$ typically present between gate layers, which may cause increased charge noise in Si/SiGe quantum dot devices \cite{Connors:2019p7549C}. Removal of grown dielectrics for electrical isolation comes at a cost though---one must rely on the AlO$_x$ that grows natively on evaporated Al, which is reported to grow between 1.6--3.0 nm by a variety of techniques \citep{Spruijtenburg:2018p143001, Quade:2000p014747, Gupta:2015p0981, Fehlner:1986p12345, Nguyen:2018p17224, Evertsson:2015p82683}. This makes high yield fabrication of devices difficult due to gate-to-gate leakage and damage from ESD.
\\ \indent In order to determine if this native oxide is sufficient to prevent leakage, we compare four different techniques used to oxidize Al gates: native oxidation (NO), thermal annealing at 250$^\circ$C (TO), plasma ashing (PA), and UV-ozone treatment (UV). All samples are natively oxidized at standard temperature and pressure before a 15 minute treatment by each oxidation method. For TO, the anneal is at 250$^\circ$C and 45\% humidity. For PA, the plasma asher used is a YES R3 Downstream Plasma Cleaner at a pressure of 5 torr with 80 sccm O$_2$ using a power of 250 W. For UV, a Samco UV-1 model is used with an oxygen flow rate of 0.5 $l\cdot$min$^{-1}$, giving an ozone concentration of 6 g$\cdot$m$^{-3}$ with its substrate platen heated to 250$^\circ$C.
\\ \indent To characterize how each oxidation treatment affects the electrical isolation properties of the Al gates, we fabricate two-layer test structures designed to replicate the overlap between plunger/barrier layers and the screening gate layer of the device shown in Figure~\ref{fig:fig1}. The test structure device design includes the portion of the plunger/barrier gate that climbs onto the screening gate lying beneath it to ensure this rugged interface was included as part of the breakdown test. The first layer is oxidized using one of the four methods, and subsequently the breakdown voltage is measured using standard current-voltage measurements. Devices are fabricated on [100] Si wafers, electrically isolated from the Si by 100 nm of ALD-grown Al$_2$O$_3$. The two layers are patterned using EBL. 30 and 50 nm of Al are deposited for the first and second layer, respectively. The structures have an overlap area of 1 $\mu$m x 0.1 $\mu$m, similar to the gate overlap in Figure~\ref{fig:fig1}(d). Each electrode layer forming the test structures is electrically connected to bond pads, as described in detail in Section \ref{interconnects}, which is used to apply a differential voltage to the device and measure leakage current using a Keithley Model 2700 Multimeter, current limited to 5 nA. Devices are tested cryogenically at 2 K by increasing the differential voltage between electrode and counter-electrode pairs until breakdown is observed, defined here to be when 100 pA of gate-to-gate leakage is measured (a measured current density of 0.1 A$\cdot$cm$^{-2}$). Devices are inspected after measurement in an SEM to ensure they have not been destroyed by ESD. 
\begin{figure}
\includegraphics[width=0.48\textwidth]{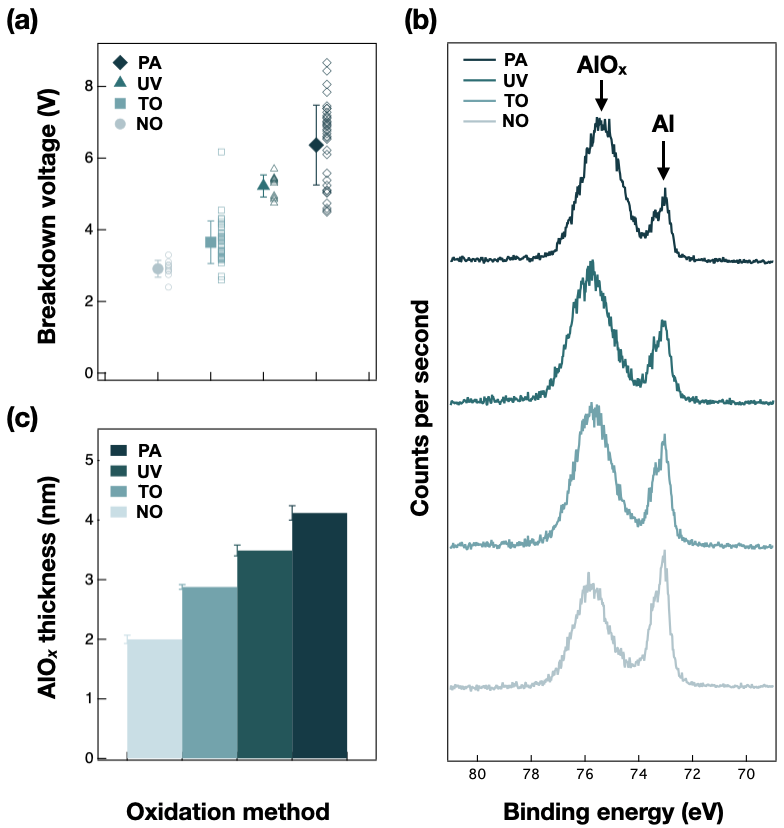}
\caption{\label{fig:fig2} Low-temperature oxidation characterization of AlO$_x$. (a) Breakdown voltage data for native (NO), thermally annealed (TO), UV-ozone (UV), and plasma ashed (PA) inter-gate AlO$_x$. Solid data points show the average breakdown voltage and standard deviation measured for each method. Adjacent to solid data points are the breakdown voltages of all devices measured for each method, shown as hollow, semi-transparent points. (b) X-ray photoelectron spectroscopy (XPS) spectra taken for each method; traces offset for clarity. The AlO$_x$ thickness is extracted from the relative intensities of the oxidic/metallic Al 2p peaks \cite{Strohmeier:1990p155156}. (c) Extracted thickness of AlO$_x$ from XPS spectra.}
\end{figure} 
\indent Figure~\ref{fig:fig2}(a) shows the results of the breakdown test. For each of the four oxidation methods, 10, 41, 11, and 35 samples are measured for NA, TO, UV, and PA, respectively, to determine an average breakdown voltage, shown as a solid data point. Adjacent to solid data points are the breakdown voltages of all devices measured for each method, shown as hollow, semi-transparent points. The average breakdown and standard deviations are summarized in Table~\ref{table:tab2}. The spread and standard deviations measured for TO and PA are a more accurate representation of their true values due to the larger number of devices measured.
\renewcommand{\arraystretch}{1.3}
\begin{table}[b]
\caption{\label{table:tab2} Electrical characterization summary of low-temperature Al oxidation methods. Oxide thickness was determined from x-ray photoelectron spectroscopy (XPS). The breakdown voltage (BDV) is defined here to be when $>$100 pA of current is measured between electrode and counter-electrode pairs.} 
\centering\begin{tabular}{c c c c c}
\hline\hline 
 &  &  &  \\[-2ex]
Method & Thickness (nm) & \hspace{13px} BDV (V) \hspace{8px} & \hspace{3px} \thead{Min. BDV (V)}  \\ [1ex] \hline	
	PA & 4.12 $\pm$ 0.12 & 6.36 $\pm$ 1.12 & 4.50 \\ 
	UV & 3.49 $\pm$ 0.09 & 5.23 $\pm$ 0.31 & 4.75 \\
	TO & 2.88 $\pm$ 0.04 & 3.65 $\pm$ 0.59 & 2.60 \\
	NO & 2.00 $\pm$ 0.07 & 2.91 $\pm$ 0.24 & 2.80 \\
	\hline\hline
\end{tabular}
\end{table}
\\ \indent The minimum breakdown voltages observed in PA and UV, shown in column four of Table~\ref{table:tab2}, are significantly above the estimated maximum differential voltage needed for overlapping Al gate devices in Si/SiGe ($\sim$1--2 V), whereas it is only modestly higher for NO and TO. As device size is increased to include more and more overlapping aluminum gates, it becomes increasingly important that the native AlO$_x$ is further oxidized to be more robust to electrical breakdown, thus PA and UV represent two low-temperature oxidation methods that mitigate failure due to gate-to-gate leakage at an acceptable level for larger devices. The distinct increase in breakdown voltage of the PA and UV techniques suggests increased AlO$_x$ thickness and/or higher quality AlO$_x$. To isolate these variables, we analyze bulk Al films using x-ray photoelectron spectroscopy (XPS) to determine oxide thickness. The XPS spectra are shown in Figure~\ref{fig:fig2}(b). The AlO$_x$ thickness can be extracted from the relative intensities of the oxidic/metallic Al 2p peaks, as detailed in Ref.~\citep{Strohmeier:1990p155156}. We note there is a slight red shift observed in the AlO$_x$ XPS peak of the PA samples which is likely due to PA being the thickest of the four films. Since the Al-O bonding energy is $\sim$0.6 eV lower than the Al-OH bonding energy \cite{Iatsunksky:2015p11352}, the surface layer of AlO$_x$ is more hydroxyl rich than bulk AlO$_x$, and the red shift occurs simply because PA is thicker than the other films. The spectra shown in Figure~\ref{fig:fig2}(b) are a subset of a larger dataset taken to determine the average thickness and fluctuations in thickness for each method. Six XPS spectra are taken at different positions on the bulk film for each oxidation method to obtain statistical fluctuations in the AlO$_x$ thickness. We note that the AlO$_x$ thickness of amorphous and crystalline oxidized Al are similar \cite{Evertsson:2015p82683}, so our bulk measurements are closely representative of the much smaller device gate electrodes, which are composed of Al grains on the order of the gate electrode width.  Figure~\ref{fig:fig2}(c) shows the results from all measured samples, and the values are summarized in Table~\ref{table:tab2}. The thickness for PA, UV and TO all exceed NO. PA displays the fastest oxidation rate. \\
\begin{figure*}[t]
\includegraphics[width=1.0\textwidth]{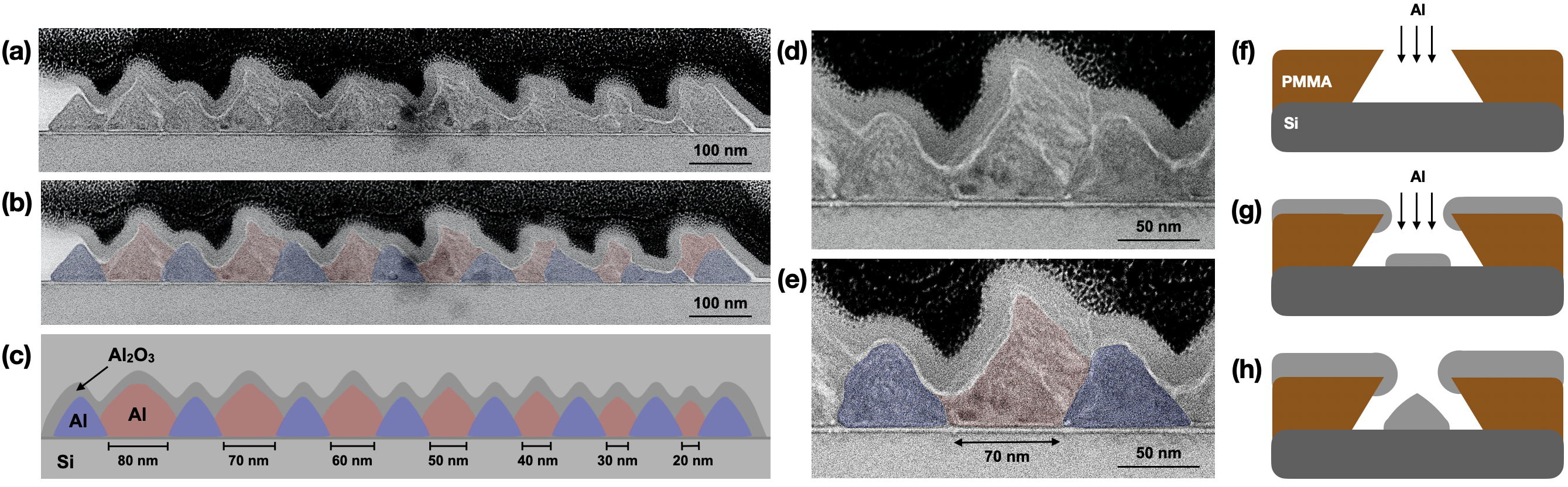}
\caption{\label{fig:fig3} Failure analysis for varying gate widths. (a, b) Scanning transmission electron microscopy (STEM) images of an overlapping Al gate structure, where (b) has been false-colored. Al plunger gates (blue) are deposited nominally 50 nm thick and 70 nm wide on a Si substrate with gate-to-gate pitches varying from 90--150 nm. Al barrier gates (red) are deposited subsequently, nominally 65 nm thick with widths varying from 40--100 nm. The test structure is capped with 20 nm of ALD-grown Al$_2$O$_3$ (gray) and a protective platinum layer (black). (c) A schematic of the expected cross-section of the overlapping Al gate structure. The sidewalls are assumed to be $\sim$60$^\circ$ as is observed on average in the STEM image. (d, e) Zoom in of Figure \ref{fig:fig3}(a), where (e) has been false-colored. The 70 nm gap between plunger gates is filled homogeneously by the barrier gate. (f-h) Schematic showing before (f), during (g) and after (h) the Al e-beam evaporation. The sloped sidewalls observed for Al gates is due to accumulation of Al on PMMA sidewalls during evaporation \citep{Spruijtenburg:2018p143001}.}
\end{figure*}
\indent Calculating a breakdown field of the AlO$_x$ layer is not as straightforward as taking the ratio of columns 3 and 2 of Table~\ref{table:tab2}, because of the complex oxidation behavior of the Al-AlO$_x$-Al stack. Nonetheless, it is useful to note that such a simple division would yield results of $\sim$12--15 MV$\cdot$cm$^{-1}$.  For context, the breakdown values reported for \emph{bulk} AlO$_x$ in literature are $\sim $3--8 MV$\cdot$cm$^{-1}$ \cite{Khosa:2018p25304, Tanner:2007p03510, Yota:2013p1A134}.  There are three reasons for the significant difference between the previously reported bulk values and the results of the ratio of columns 3 and 2 of Table~\ref{table:tab2}.  First, column 2 reports the measured thickness of films that did not have Al evaporated on top of the oxide.  The bottom part of Al electrodes has been shown to oxidize when in contact with oxygen-rich materials such as SiO$_2$ \cite{Lim:2009:p173502} and ALD-Al$_2$O$_3$ \cite{Brauns:2018p24004}. Thus, the AlO$_x$ thickness between the two Al electrodes may increase upon deposition of the top Al electrode, presumably with a less oxygen-rich composition.  Second, the AlO$_x$ in these samples is in the ultrathin regime, and such films display an enhanced breakdown field \cite{Lin:2005p82904}.  Third, the substrate temperature is 2 K rather than room temperature for these measurements, and such a decrease in temperature can increase breakdown fields \cite{Wu:2007p72105, Yota:2013p1A134}.  All three of these mechanisms are likely to contribute to the large ratio of columns 3 to 2 in Table~\ref{table:tab2}. \\
\indent The oxidation rate of Al for both UV and PA follows a $d \sim \sqrt{t}$ dependence \citep{Gupta:2015p0981, Quade:2000p014747}, where $d$ is thickness and $t$ is time, which can be used to further increase the AlO$_x$ thickness if desired. PA is implemented in the fabrication of the triple-dot device shown in Figure~\ref{fig:fig1}. Low-frequency charge noise was measured in two devices using PA on separate chips, using the technique detailed in Ref.\ \cite{Freeman:2016p253108}, obtaining values of 2.31 and 0.89 $\mu \text{eV}/\sqrt{\text{Hz}}$ at 1 Hz where the noise power spectral density follows a $1/f$ dependence with exponents 1.03 and 1.04, respectively. This is consistent with other recent results for Si/SiGe quantum dots using ALD-grown Al$_2$O$_3$ as a gate dielectric \citep{Connors:2019p7549C, Freeman:2016p253108}.
\section{STEM Failure Analysis}
\indent To further investigate potential failure modes in overlapping Al gate devices, we fabricate test structures using PA between gate layers and take cross-sectional STEM images (Figure~\ref{fig:fig3}). Two-layer test structures are fabricated to investigate filling of the gaps between plunger gates (first layer) by barrier gates (second layer) for varying gate widths. 
\\ \indent The test structures are fabricated using the same procedure described for the device shown in Figure~\ref{fig:fig1}(a-d), except the first layer is omitted. We note that the evaporator hearth is water-cooled and copper radiation shielding is used to keep the sample stage near room temperature to reduce high-temperature induced morphological effects such as large grain size and sloped sidewalls of the aluminum \cite{Muller:2015p39036}. The gate geometry is shown in Figure~\ref{fig:fig3}(a, b), where the plunger gate layer consists of eight 70 nm wide, 50 nm thick gate electrodes with gate-to-gate pitches varying from 90--150 nm in steps of 10 nm (increasing right to left). This leaves nominal gaps 20--80 nm wide for the barrier gates to fill. The barrier gates are evaporated 20 nm wider than each gap, ranging from 40--100 nm (increasing right to left) and the thickness of the barrier gate layer is nominally 65 nm. A schematic of the expected cross-section is shown in Figure~\ref{fig:fig3}(c), where the sloped sidewalls and the effect on gate morphology has been taken into account. This schematic can be compared to the region of interest of the triple-dot device, which is shown earlier in Figure~\ref{fig:fig1}(e). In Figure~\ref{fig:fig3}(b), the plunger gate layer (blue) and barrier gate layer (red) have been false-colored using the dark-field STEM image shown in the Supplemental Material.
\\ \indent Several striking features are revealed from the STEM images. The sidewall slope of the plunger and barrier gate electrodes is found to be between 45--60$^\circ$. This is consistent with AFM profiling shown in Ref.\ \cite{Muller:2015p39036} and can be explained by the process illustrated in Figure~\ref{fig:fig3}(f-h). During evaporation, Al accumulation narrows the opening of the PMMA mask. In extreme cases, the opening closes completely, leaving gate electrodes thinner than intended. This symptom is visible on the far right of Figure~\ref{fig:fig3}(a, b), and can lead to yield problems associated with step coverage failure.
\\ \indent On the far right of Figure~\ref{fig:fig3}(a, b), the plunger gates themselves are deformed and reduced in thickness. This is attributed to resist wall collapse \cite{Nealey:200p018913}, since the thickness is skewed to one side. Top-down SEM images (not shown) are consistent with such resist collapse.  Resist wall collapse is an issue in these devices because the barrier gates are relatively thick (65 nm), to guarantee high-yield connection over the many underlying gates in this device.  As a consequence, the PMMA resist is also chosen to be thick (180 nm), for high-yield liftoff.
\\ \indent The barrier gates fill and contact the underlying surface across all regimes, independent of the gap width. The only case in this test structure where a barrier gate is not in contact with the substrate is on the far right side, where the intended gap of 20 nm completely closed off. In the narrower regime, the barrier gate layer fails for a different reason than the plunger gate layer. The barriers do not hold their intended thickness and shape, due to the process illustrated in Figure~\ref{fig:fig3}(f-h). Given the  45--60$^\circ$ sloped sidewalls, the thickness of the gate cannot reliably exceed the width using this recipe. A way to design around this is to pattern barrier gates as wide as possible when devices contain small gaps between plunger gates. 

\section{Device Interconnects}
\label{interconnects}
\begin{figure}
\includegraphics[width=0.48\textwidth]{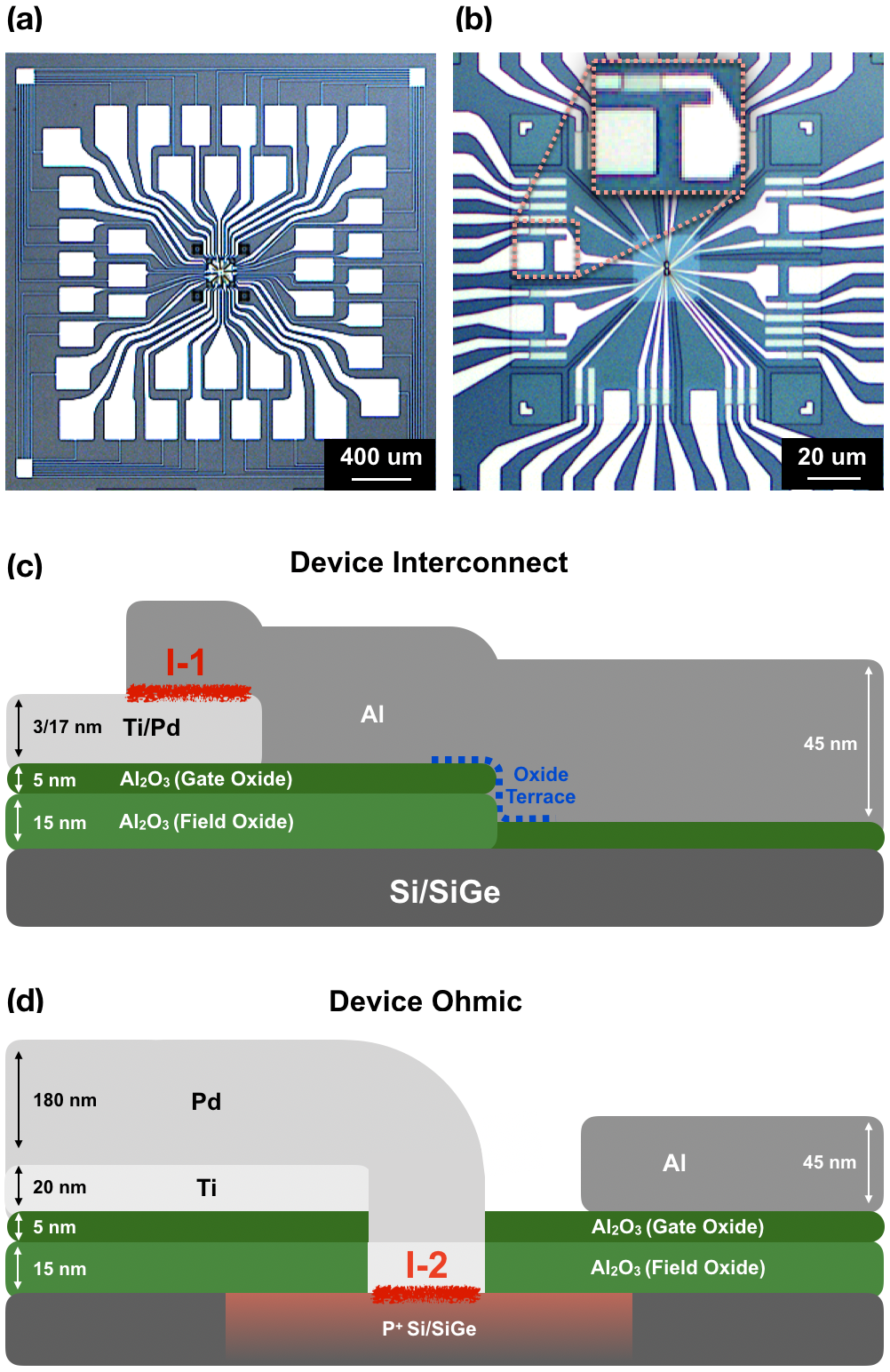}
\caption{\label{fig:fig4} Interconnect and ESD protection architecture. (a) Optical image of full device. The electrostatic discharge (ESD) protection wiring borders the device, shorting all gate leads together during fabrication of the device. (b) Optical image of the device mesa. Interconnects between the active region and bond pads are joined by 3/17 nm thick Ti/Pd pads. (Inset) Top-down zoom-in of a Pd-Al interconnect and ohmic, corresponding to the cross-sectional schematic shown in (d). (c) Schematic cross-section (not to scale) of a device interconnect where the Al gate lead meets the Pd interconnect. An oxide terrace is used to reduce bulk oxide beneath the active region. At I-1, an interface between Al and Pd is formed, resulting in problematic alloys at $\sim$300$^\circ$C. (d) Schematic cross-section (not to scale) of a device interconnect at the ohmic site. An interface (I-2) between Ti/Pd and the P$^+$ doped Si/SiGe heterostructure forms. Diffusion processes are observed to begin at $\sim$400$^\circ$C, which can affect the interconnect morphology and electrical connectivity.}
\end{figure}

\indent Another challenge in fabricating quantum dots in Si/SiGe heterostructures is the design considerations needed for developing on-chip interconnects between bond pads and the active region. High temperature processes are desirable at different stages of device fabrication, but often times they cause failure of interconnects. Below, we discuss critical interfaces affecting the thermal budget at various stages of fabrication, how to design around the thermal budget, and integration of on-chip ESD protection.
\\ \indent For the interconnect design, we etch a mesa between the bond pads and active region, shown in Figure~\ref{fig:fig4}(b), preventing gate-to-ohmic leakage in the event that wire bonds punch through the substrate into the 2DEG. The yield of devices can be severely limited by damage due to ESD after the fabrication of the active region. To ameliorate this, on-chip ESD protection is implemented. Shorting wires are patterned before fabrication of the active region in the same lithographic step as bond pads, preventing build-up of charge between gates. The shorting wires can be seen Figure~\ref{fig:fig4}(a) bordering the bond pads. An equipotential for all gates is maintained through device packaging by wire bonding, grounding through a printed circuit board, and physically scribing away the shorting wires on-chip afterwards. Alternatively, the leads can be electrically disconnected after fabrication using an etch step; however, this does not protect the device during packaging.  
\\ \indent Consideration of the thermal budget is important when determining a process flow for quantum dot devices. Significant interdiffusion of the Si/SiGe interface at the quantum well can occur above 800$^{\circ}$C \cite{Xia2007p:30904}, lowering the valley splitting \cite{Friesen:2006p202106}. This makes growth of high quality SiO$_2$ \cite{Irene:1978p31277, Yoneda:1995p45802} on Si/SiGe heterostructures difficult, so instead ALD is used to grow amorphous Al$_2$O$_3$ or HfO$_2$ for gate-to-2DEG isolation. After the field and gate oxide is grown, thermal annealing can be used to reduce interface trapped charge \cite{Wolf:2000, Spruijtenburg:2016p38127}, which has been shown to reduce threshold voltages and increase transconductance in Si-MOS quantum dots \cite{Angus:2007p845}. \\ 
\indent The presence of interfaces (Figure~\ref{fig:fig4}(c), I-1 and Figure~\ref{fig:fig4}(d), I-2) further limits the thermal budget of devices. We also note that the presence of thin Al films imposes its own set of thermal constraints, including dewetting and void formation, occurring at 400$^\circ$C \cite{Spruijtenburg:2018p143001} and between 300--500$^\circ$C \cite{Hinode:1989p23447}, respectively. To maximize the thermal budget of devices, we choose palladium instead of gold for interconnect metallization. Au is often used, but formation of AuAl$_2$ (purple plague), a non-conductive alloy, is observed in thin Au-Al films at temperatures as low as 217$^\circ$C on a minutes time scale \cite{Galli:1980p1155}. Additionally, at the ohmic metallization site, the Au/Si interface experiences an interstitial diffusion process that affects thin film electrode morphology at temperatures as low as 200$^\circ$C \cite{Poate:1981p07579}. We observe a similar process when annealing a Au/Ti/Si structure analogous to I-2 at 250$^\circ$C for 30 minutes. When using Pd, the thermal budget is found to increase. Formation of Pd-Al alloys in thin films near 300$^\circ$C \cite{Koster:1978p8035} can cause electrical discontinuities, which we verified with a 30 minute anneal at 300$^\circ$C on a 30/17/3 nm Al/Pd/Ti gate stack. Additionally, we observe diffusion processes at 400$^\circ$C occurring at the Pd/Ti/Si interface, affecting ohmic gate morphology. The consequence of these critical temperatures is that they cannot be exceeded by processes such as post-metallization anneals or inter-gate oxidations with the expectation of high yield. However, the UV and PA methods presented in this paper do not exceed any of these critical temperatures when using Pd as an interconnect/ohmic gate metal (Figure~\ref{fig:fig4}(b), inset), and thus are good choices for oxidation of inter-gate oxide. 

\section{Conclusion}
\indent We have demonstrated an improved fabrication process and performed failure analysis on many critical interfaces in overlapping Al gate quantum dot devices. The main takeaway from the AlO$_x$ characterization is that PA and UV mitigate failure due to gate-to-gate leakage and ESD, and their process temperatures are compatible with the thermal budget of overlapping Al gate quantum dot devices. Optimization of individual oxidation techniques, not considered here, may improve oxide quality and allow further tuning of the AlO$_x$ thickness. For UV oxidation of Al, relative humidity, time of oxidation, temperature, and partial pressure of oxygen all affect the growth rate \citep{Gupta:2015p0981, Fehlner:1986p12345, Ramanathan:2003p6416}. For PA, time of oxidation, excitation frequency, power, and partial pressure of oxygen affect the growth rate \citep{Quade:2000p014747, heij:2019p12345}.
\\ \indent We also fabricated overlapping Al gate two-layer structures and used cross-sectional STEM images to analyze failure modes of varying gate geometries. The sidewalls of the plungers/barriers were found to be between 45--60$^\circ$, creating potential issues with step coverage near the active region. This can often be designed around by maximizing the barrier gate widths when devices have small gaps between plunger gates. We also identified failure modes associated with critical dimensions in device geometries. Gates with $<$40 nm gaps between plungers showed abnormalities due to resist wall collapse and thinner than intended deposition thickness. 
\\ \indent Finally, we showed an interconnect fabrication process that implements on-chip ESD protection and identified critical temperatures that can result in electrical discontinuities at different stages of device fabrication. By designing a fabrication process compatible with these temperatures, field/gate oxides may still be annealed at 450$^\circ$C in forming gas before metal deposition, and inter-gate oxide in Al devices can be further oxidized using UV or PA. The result of a process with these changes implemented was shown in Figure~\ref{fig:fig1}(e, f), where the quantum dots were notably stable and all gate electrodes worked as intended. 

\begin{acknowledgments}
We thank Jason Petta and Lieven Vandersypen for helpful discussions. Research was sponsored in part by the Army Research Office (ARO) under Grant Numbers W911NF-17-1-0274 and by the Vannevar Bush Faculty Fellowship program under ONR grant number N00014-15-1-0029. We acknowledge the use of clean room facilities supported by NSF through the UW-Madison MRSEC (DMR-1720415) and electron beam lithography supported by the NSF MRI program (DMR-1625348). The views and conclusions contained in this document are those of the authors and should not be interpreted as representing the official policies, either expressed or implied, of the Army Research Office (ARO), or the U.S. Government. The U.S. Government is authorized to reproduce and distribute reprints for Government purposes notwithstanding any copyright notation herein.
\end{acknowledgments}

\appendix*
\section{Detailed Fabrication Process}
\label{sup:appendix}

The fabrication process flow for devices integrating low-temperature oxidation of Al gates, on-chip ESD protection, thermal annealing of the field/gate oxide, and the use of Pd as interconnect metal is discussed below. Patterning for all photolithography steps is done using a Nikon NSR-2005i8A i-line stepper. All photolithography steps precede EBL steps and are completed on a 3" Si/SiGe wafer before it is diced into chips for EBL. A base mesa structure, shown in Figure~\ref{fig:fig4}(a, b), accommodates a range of device geometries with up to 40 gates. All Al$_2$O$_3$ etch steps use 20:1 BOE as an etchant.  
\\ \indent Device fabrication begins with a mesa etch using reactive ion etching (RIE) with CHF$_3$ as the process gas. Before ion implantation, a 10 nm screening oxide (Al$_2$O$_3$) is grown using ALD. This helps prevent cross-linked resist from contaminating the substrate when stripping the resist mask post-ion implantation. $^{31}$P$^+$ is implanted at 25 keV with a density of 5$\times$10$^{15}$ ions$\cdot$cm$^{-2}$ at a 7$^{\circ}$ tilt to electrically connect to a 40 nm deep Si quantum well. The implanted region is then annealed at 700$^\circ$C for 15 seconds in forming gas using a rapid thermal annealer (RTA). 15 nm of field oxide and 5 nm of gate oxide (Al$_2$O$_3$) are grown subsequently using ALD. Between oxide depositions, a 15 $\mu$m $\times$ 15 $\mu$m terrace is etched in the field oxide to reduce bulk oxide in the active region. Each layer is annealed individually using a forming gas anneal (FGA) at 450$^\circ$C. Two anneals, one after each layer, is preferred over annealing both at the same time due to blistering of ALD Al$_2$O$_3$ \cite{Beldarrain:2013p68170, Vermang:2011p03567}.
\\ \indent Metallization of devices begins with a two step process to etch and metallize the ohmic contacts. Positive photoresist with an HMDS adhesion layer is used to prevent undercutting during the etch and a negative resist is used for ohmic metallization. Interconnect jumper pads are then deposited using a 3/17 nm Ti/Pd metal stack. This total thickness must be sufficiently thin compared to the first Al gate layer thickness to prevent step coverage failure. Bond pads and the ESD protection wiring are deposited together using a 20/180 nm Ti/Pd metal stack. 
\\ \indent After wafer-level photolithography steps have been completed, the wafer is diced and the active region is fabricated on chips. Three layers of Al gates are deposited using an e-beam evaporator with thicknesses of 30/50/65 nm, increasing from first to last. After the first and second layers are deposited, the native AlO$_x$ that grows on the gate electrodes is further oxidized as discussed in the main text.

\bibliography{main.bib}

\end{document}


\def\simlt{\mathrel{\lower .3ex \rlap{$\sim$}\raise .5ex \hbox{$<$}}}

\title{\textbf{\fontfamily{phv}\selectfont 
Supplemental Material: \\ Fabrication process and failure analysis for robust quantum dots in silicon}}
\author{J. P. Dodson}
\affiliation{Department of Physics, University of Wisconsin-Madison, Madison, WI 53706, USA}
\author{Nathan Holman}
\affiliation{Department of Physics, University of Wisconsin-Madison, Madison, WI 53706, USA}
\author{Brandur Thorgrimsson}
\affiliation{Department of Physics, University of Wisconsin-Madison, Madison, WI 53706, USA}
\author{Samuel F. Neyens}
\affiliation{Department of Physics, University of Wisconsin-Madison, Madison, WI 53706, USA}
\author{E. R. MacQuarrie}
\affiliation{Department of Physics, University of Wisconsin-Madison, Madison, WI 53706, USA}
\author{Thomas McJunkin}
\affiliation{Department of Physics, University of Wisconsin-Madison, Madison, WI 53706, USA}
\author{Ryan H. Foote}
\affiliation{Department of Physics, University of Wisconsin-Madison, Madison, WI 53706, USA}
\author{L. F. Edge}
\affiliation{HRL Laboratories, LLC, 3011 Malibu Canyon Road, Malibu, CA 90265, USA}
\author{S. N. Coppersmith}
\affiliation{Department of Physics, University of Wisconsin-Madison, Madison, WI 53706, USA}
\affiliation{University of New South Wales, Sydney, Australia}|
\author{M. A. Eriksson}
\affiliation{Department of Physics, University of Wisconsin-Madison, Madison, WI 53706, USA}

\maketitle
\onecolumngrid

\begin{figure*}[h]
\includegraphics[width=0.95\textwidth]{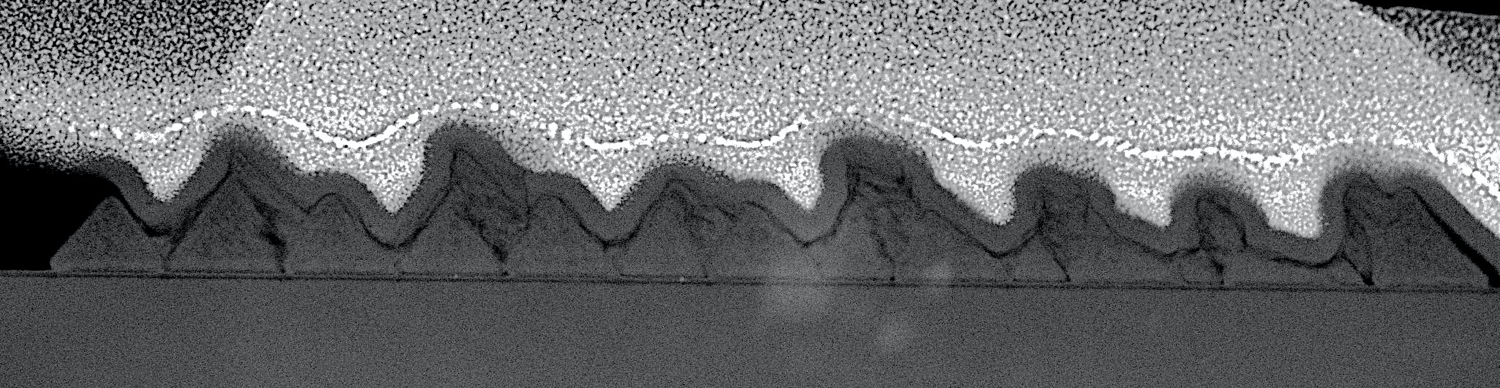}
\caption{\label{fig:figS1} STEM image of overlapping aluminum gate test structure using the high-angle annular dark-field (HAADF) detector. This high contrast image was used to aid in identifying interfaces between plungers and barriers and false-coloring the plunger and barrier gate layers shown in the main text.}
\end{figure*}
